\def\etal{{et al.\ }}

\def\gxs{galaxies}
\def\U3{U_{300}}
\def\B4{B_{450}}
\def\V6{V_{606}}
\def\I8{I_{814}}
\documentclass{aa}
\authorrunning{M. Volonteri}
\usepackage{psfig}
\usepackage{amssymb}
\begin{document}

   \thesaurus{06     
              (03.11.1;  
               16.06.1;  
               19.06.1;  
               19.37.1;  
               19.53.1;  
               19.63.1)} 
   \title{A Catalogue of Galaxies in the HDF-South: Photometry and Structural Parameters\thanks{The full catalogue is available in electronic form at the CDS via anonymous ftp to cdsarc.u-strasbg.fr (130.79.128.5) or via http://cdsweb.u-strsbg.fr/Abstract.html}}


   \author{ M. Volonteri$^2$, P. Saracco$^1$, G. Chincarini$^{1,2}$
}

   \offprints{Marta Volonteri (marta@merate.mi.astro.it)}

 \institute{$^1$ Osservatorio Astronomico di Brera-Merate, Italy, $^2$ Dipartimento di Fisica, Univ. di Milano, Italy} 
   \date{Received ......, 2000; accepted }

   \maketitle

   \begin{abstract}We describe the construction of a catalogue of galaxies in the optical field of the Hubble Deep Field South.  The HDF-S observations produced WFPC2 images in U, B, V, and I, the version 1 data have been made public on 23 November 1999. The effective field of view is 4.38 arcmin$^2$ , and the 5$\sigma$ limiting magnitudes (in a FWHM aperture) are 28.87, 29.71, 30.19, 29.58 in the U, B, V and I bands, respectively. We created a catalogue for each pass-band (I$_{814}$, V$_{606}$, B$_{450}$, U$_{300}$), using simulations to account for incompleteness and spurious sources contamination. Along with photometry in all bands, we determined on the I$_{814}$-selected catalogue (I$_{814}<26$) structural parameters, such as a metric apparent size, derived from the petrosian radius, an asymmetry index,  light concentration indexes and the mean surface brightness within the petrosian radius. 

               
   \end{abstract}

%

\section{Introduction} 
The Hubble Deep Field South (HDF-S) was observed in October 1998 by the Hubble Space Telescope (HST). It is the southern counterpart of the Hubble Deep Field North (HDF-N) and shares its characteristics of depth and spatial resolution. The HDF-S is a four arcmin$^2$ survey located at RA 22$h$ 32$m$ 56$s$, DEC -60$^\circ$ 33' 02'', observed during 150 orbits with the Wide Field Planetary Camera 2 (WFPC2). The WFPC2 detector is composed by 4 chips: the Planetary Camera (PC) with higher resolution  (0.05 arcsec/pixel) and smaller field of view (35$\times$35 arcsec$^2$) and 3 Wide Field Cameras (WF), with a spatial resolution of 0.10 arcsec/pixel and a  field of view of 77$\times$77 arcsec$^2$.We will refer to the area of sky observed by the Planetary Camera as PC field, and to the one observed by the three Wide Field Cameras as WF field. 

HDF-S images cover a wavelength range from the ultraviolet to near-infrared (4 broad-band filters roughly corresponding to standard UBVI). Images were taken in four filters: F300W, F450W, F606W, and F814W, by a dithering technique. Released images are a combination of all individual exposures, weighted for the background signal, and resampled to a pixel scale of 0.0398 arcsec/pixel using the "Drizzle" package (Fruchter, Hook \etal 1997). Details about observations and data reduction may be found in the Hubble Deep Field South web page (http:// www.stsci.edu/ftp/science/hdfsouth/hdfs.html).

The main goal of the two HDF campaigns is to study the characteristics of faint galaxies and provide constraints on models explaining the formation and evolution of galaxies, in particular the excess of faint blue sources. It is therefore crucial to have a reliable catalogue, both for sources detection and photometry. In this paper we present a galaxy catalogue derived from the public version 1 images in fits format of the WFPC2 data, released on 23 November 1999, along with documentation about data reduction and absolute calibration in the different bands. 

This catalogue gives for each galaxy, besides photometric information in all the optical bands, metric size, mean surface brightness, asymmetry index and light concentration indexes, all fundamental for different cosmological and evolutionary tests.

In Section 2 we describe the applied detection procedure and in Section 3 the technique used to compute magnitudes, while  in Section 4 we deal with the selection criteria adopted to extract the sample. In Section 5 we summarize the estimation of the sample completeness, while we discuss the possible oversampling of sub-galactic structures, such as HII regions, in high resolution imaging in Section 6. Finally in Section 7 and 8 we describe the technique used to recover galaxy colours and structural parameters respectively.
\section{Sources Detection}
 Object detection was performed across the  area 
observed in optical bands (WFPC2 field) using the software for automatic sources detection SExtractor (Bertin and 
Arnouts 1996). The software firsts smoothes the images and then applies a threshold to identify peaks. As a smoothing function we have chosen a Gaussian function having FWHM equal to the one measured on the images (i.e. 0.16 arcsec) and a detection threshold of 1 sigma per pixel, with  a minimum detection area equal to the seeing disk, have been adopted to pick up objects. The software allows to examine every identified group of contiguous pixels to deblend overlapping sources: two sources belonging to the same group are considered different objects if they differ of 5 magnitudes or less (\emph{deblend-mincont}=0.01, \emph{deblend-ntresh}=32).

This choice of low thresholds aimed at obtaining a raw catalogue not biased against very faint or small source, in fact the first sample consisted of  6093, 4747, 9850, 5229 detections in the F300W, F450W, F606W and F814W band respectively. From this sample we then extracted a catalogue optimizing completeness while minimizing the number of spurious detections by selection criteria based on the signal-to-noise ratio of sources and simulations (as explained in Section 4). 

\section{Photometry}
Galaxies in this survey span a wide range both in redshift and in physical size. A fixed aperture would measure different fractions of flux for galaxies dissimilar in shape and size and at different redshifts, while isophotal magnitudes suffer from the $(1+z)^5$ cosmological dimming in surface brightness.

``Pseudo-total'' magnitudes were therefore estimated  using the method 
of Djorgovski et al. (1995) and Smail et al. (1995): they assigned the isophotal magnitude to sources with isophotal diameter larger than  $\theta_1\approx2-3$ FWHM, while smaller sources are assigned an aperture corrected magnitude, that is the magnitude within an aperture $\theta_1$ corrected to the  magnitude within $\theta_2$, larger than $\theta_1$: $m=m(\theta1)+\Delta m $, $\Delta m=<m(\theta2)-m(\theta1)>$. In literature the choice of $\theta_1$ and $\theta_2$ is based on multiples of the FWHM of images. In the HDF-S the excellent seeing needs a different approach, we therefore estimated magnitudes for our sample on the basis of the following steps: 
\begin{itemize}
\item  For large sources, i.e. those sources having an isophotal 
diameter $D_{iso}>\theta_1$, and for ``blended'' sources, as flagged by SExtractor, we selected the SExtractor isophotal corrected magnitude;
\item to small sources, having $D_{iso}<\theta_1$, we assigned an  aperture corrected magnitude (estimated within $\theta_1$ and
then corrected by $\Delta m$ to $\theta_2$, being $\theta_1<\theta_2$) 

\end{itemize}

$\theta_1$ has been defined as the minimum apparent diameter of a galaxy having an effective diameter $r_e=10$ Kpc. Hereafter we use a $\Lambda=0$ cosmology, with $q_0=0.5$ and $H_0=50$ kms$^{-1}$Mpc$^{-1}$ unless  differently specified.
With this choice $\theta_1=1.2$ arcsec. Since the correction  $\Delta m$ is measured on a subsample of relatively bright galaxies, we  defined an area for each band, $A_{90}$, such that 90$\%$ of galaxies belonging to the \emph{djorg} subsample has isophotal area smaller than $A_{90}$. $\theta_2$ is defined as the diameter corresponding to a circle of area  $A_{90}$.

At bright magnitudes Kron's technique, based on an adaptive aperture, $r_{1}=\frac{\sum rI(r)} {\sum I(r)}$, gives very good results.  Kron (1980) and Bertin \& Arnouts (1996) demonstrated that a  photometry within an adaptive aperture ($2.5r_{1}$) is expected to measure a fraction of the total flux between 0.9 and 0.94. We then chose to use Kron's magnitude (m$_{kr}$) as a reference in order to test our method: if the flux fraction measured by Kron's technique is 0.94, the fraction estimated by our ``pseudo-total'' magnitude (m$_{dj}$) is $ x=0.94 \cdot 10^{0.4(m_{kr}-m_{dj})} $. Our procedure is intended to correct the systematic underestimate (6-10$\%$) of total flux typical of Kron's technique, which may be important for very faint sources.

 Table 1 and Figure 1 show clearly that statistically $\Delta m$ corrects for the  flux underestimate typical of Kron's magnitudes. Moreover our estimate of total magnitudes has a narrower distribution at low S/N than Kron magnitude, see Figure 3, and in a plot magnitude-isophotal area (Figure 2) it is not evident any discontinuity  in the passage between large and small sources (i.e. sources with $\theta>\theta_1$ or vice versa). This test confirms the validity of our choice of $\theta_1$.  

If not differently specified, magnitudes are expressed in the AB system, that is a system based  on a spectrum which is flat in $f_\nu$: m=$-2.5 \log f_\nu -48.60$ (Oke 1974) .

\begin{table}
\caption{Residuals m$_{kr}$-m$_{dj}$ vs. m$_{kr}$ for the V$_{606}$ selected sample, if $\theta_1=1.20$ arcsec and $\theta_2$ defined as a function of $A_{90}$. }
\begin{tabular}{ccccc}
\\ 
\hline
Filter & $\theta_2$ (arcsec) &$\Delta$m &   med(m$_{kr}$-m$_{dj}$) & x\\
\hline
F300W  &1.59 & 0.132  &     0.08$\pm$0.40 &   1.01\\
F450W  &1.85 & 0.163  &    0.11$\pm$0.15  &   1.04\\
F606W  &2.15 & 0.080 &    0.06$\pm$0.10  &   0.99 \\
F814W  &2.01 & 0.145  &    0.12$\pm$0.12  &   1.05   \\ 
\hline
\end{tabular}
\end{table} 
\begin{figure}
\centerline{\psfig{figure=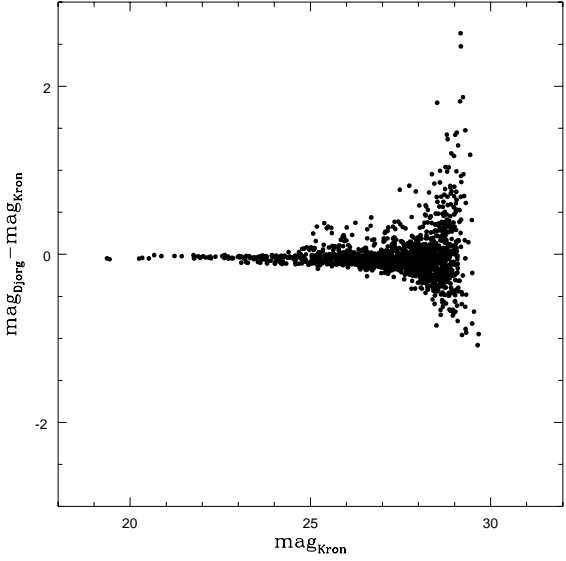,height=80mm}}
\caption{Residuals m$_{kr}$-m$_{dj}$ vs. m$_{kr}$ for the V$_{606}$ selected sample, if $\theta_1=1.20$ arcsec, $\theta_2$ as a function of  $A_{90}$.}
\end{figure}

\begin{figure}
\centerline{\psfig{figure=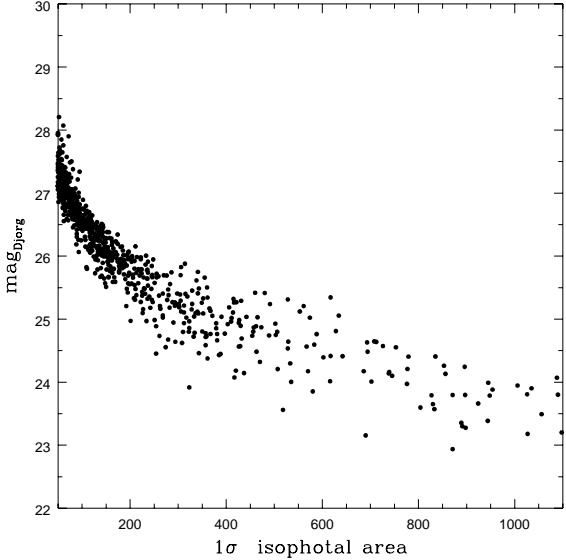,height=80mm}}
\caption{Plot of \emph{pseudo-total} magnitude vs isophotal area: it is not evident any discontinuity  in the passage between large and small sources (i.e. sources with $\theta>\theta_1$, corresponding to isophotal area $>706$ pixel$^2$). This confirms our choice of $\theta_1$.}
\end{figure}

\begin{figure}
\centerline{\psfig{figure=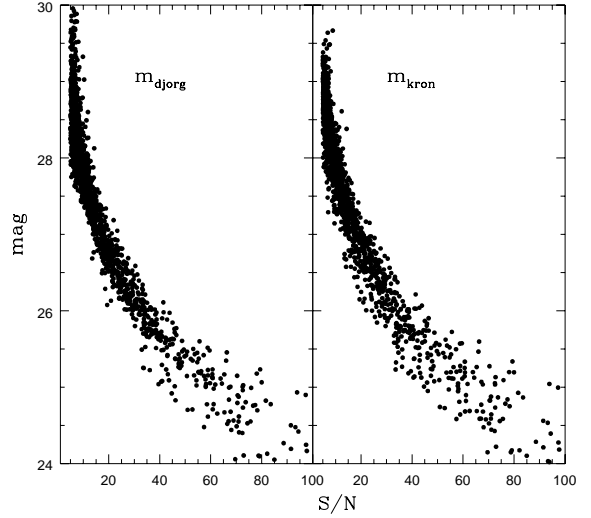,height=80mm}}
\caption{Kron's magnitude and our \emph{pseudototal} magnitude vs S/N. At low S/N \emph{pseudototal} magnitude has a narrower distribution than Kron's magnitude. }
\end{figure}

\section{Selection of the Catalogue}
Our low detection threshold led to a large number of detections (more than 4500 on less than 5 arcmin$^2$): it is therefore necessary to evaluate the number of spurious sources. The depth and coverage for each filter is not homogeneous in the field of view, due to the variety of pointings that were combined together. The image depth fades towards the edges of the area covered, as well as in a cross-shaped area between detectors which received much lower coverage than the central region of each chip. As a consequence of the decreased image quality, the outer regions of each image are less reliable: in fact sky RMS is higher than the average one and some sources near the edges revealed at inspection to be spurious. We selected reliable sources by means of criteria based on S/N ratio and comparisons with simulations.

The "Drizzle" algorithm (Variable-Pixel Linear Reconstruction) used to combine the various pointings preserves photometry and resolution and removes the effects of geometric distortion, but it causes adjacent pixels to be correlated. The pixel-to-pixel noise ($\sigma_{sky}$)therefore underestimates the true noise of a larger area by a factor  1.9. The noise measured on PC field is greater by a factor $\approx2$ than that measured on the WF area. The S/N is then computed by a semi-empirical model (Pozzetti et al. 1998, Williams et al. 1996): S/N=R/$\sigma_{tot}$, where R are net counts and $\sigma_{tot}^2=R/(\Gamma t_{exp})+2\cdot1.9^2\sigma_{sky}^2A_{obj}$ for WF sources, $\sigma_{tot}^2=R/(\Gamma t_{exp})+2\cdot 4\cdot1.9\sigma_{sky}^2A_{obj}$ for PC sources. In the above formulas $\sigma_{sky}$ is the pixel-to-pixel sky RMS,  t$_{exp}$ is the exposure time,  $\Gamma$ is the gain expressed in electrons per ADU,  A$_{obj}$ and A$_{sky}$ are respectively the object and the sky isophotal areas (in pixel) used to estimate the local background.

The former term in the sum represents the Poissonian noise due to the source, the latter estimates statistical fluctuations in the mean value of sky, in the Poissonian approximation. The factor 2 is linked to uncertainties in the determination of local background: the correct term would be  $1.9^2\Gamma^2\sigma_{sky}^2A_{obj}^2/A_{sky}$, but since A$_{sky}$  differs less than 30$\%$ by the mean value of A$_{obj}$, we considered A$_{sky}\approx$A$_{obj}$.

Our detection threshold corresponds to a minimum signal-to-noise ratio 
of $S/N_{WF}=1.34$ and $S/N_{PC}=0.67$ for the faintest sources
detectable on the WF area and on the PC area respectively.

In Table 2 we report for each filter the zeropoint (AB magnitude, Oke 1974),
the sky RMS estimated by SExtractor and the corresponding
5$\sigma$ magnitude limit for a point source.

\begin{table}
\caption{RMS sky values and 5$\sigma$ magnitude limit for
the different bands.}
\begin{center}
\begin{tabular}{llll} 
\hline
Filter& zeropoint& RMS&m$_{lim}$  \\
 &   &  (ADU/pix)$\times10^{-5}$& \\
\hline
F300W & 20.77&	1.674& 28.87\\
F450W& 21.94&	2.284&	29.71\\
F606W&	23.04&	4.126&	30.16\\
F814W&	22.09&	2.960&	29.58\\
\hline
\hline
\end{tabular} 
\end{center}
\end{table}

We treated this problem statistically, in the hypothesis that noise is symmetrical with respect to the mean sky value. Operationally we have first created for each filter a noise frame by reversing the original images, in order to reveal the negative fluctuations and to make negative (i.e. undetectable) real sources (Saracco et al. 1999).
Then we run SExtractor with the same detection parameter set used to search for sources in the original images detecting, by definition, only spurious sources. Applying  a S/N=5 cut off, after removing the edges of the images, we were able to reduce the spurious contamination to a negligible fraction (4$\%$) on the WF area, while  such a cut off is not able to  reduce  spurious detections to a reasonable level on the PC area  being them more than 35$\%$. 
In Figure 4  the magnitude distribution of spurious sources obtained on the WF area and the PC area in the F606W band are shown.  It is clear that the influence of spurious sources on the PC field is still remarkable after applying selection criteria, while the contamination is suppressed in the WF field.

Thus, to avoid introducing such a large number of spurious by the PC data,  we restricted the selection of sources to the central WF area only corresponding to 4.38 arcmin$^2$. On this area 450, 1153, 1694 and 1416 sources have been selected accordingly to the above criteria in the F300W, F450W, F606W and F814W band respectively, while the raw catalogues had  6093,4747, 9850, 5229 detections in the same bands.

In every magnitude bin we compared the number of sources in our final catalogue with the number of spurious detections in order to get the contamination of false detection, shown in Table 3.

\begin{table}
\caption{ Percentage of spurious detections in every magnitude bin, as estimated comparing detections on the on reversed (i.e. multiplied by -1) images with detections on real frames, that is percentage ratio of objects detected on reversed frames to sources detected in original frames in every magnitude bin.}
\begin{tabular}{ccccccc}   
\\ \hline
Pass-band & 26.25& 26.75  & 27.25& 27.75 & 28.25& 28.75  \\  
\hline
U$_{300}$       & 1  &5.8  & - & - &-&-\\
B$_{450}$       &  0 &  0  & 0.9&1.9& 2.9&-\\
V$_{606}$       &  0 &  0  & 0&0.3& 1&3\\
I$_{814}$       &  0 &  0.5& 0.8& 4.2&-&- \\
\hline
\end{tabular}
\end{table}
\begin{figure}
\centerline{\psfig{figure=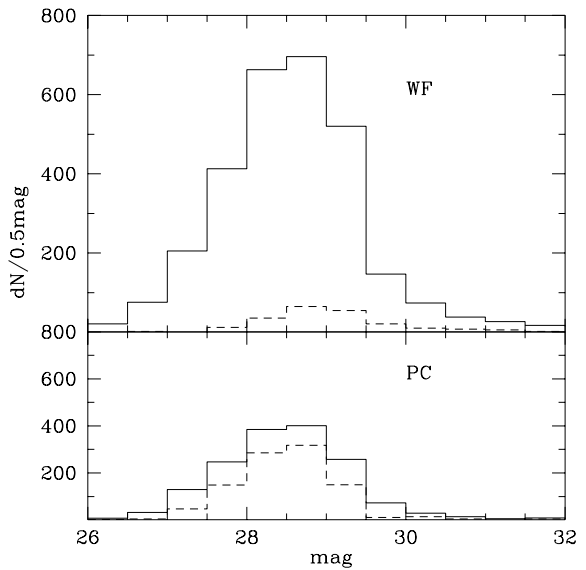,height=80mm}}
\caption{Spurious sources, defined as detections on the reversed (i.e. multiplied by -1) V$_{606}$ image.  The upper panel refers to the WF field, the lower panel to the PC field. The solid line represents the initial detections, the dashed line represents spurious sources left after applying our selection criteria.}
\end{figure}
We then removed stars from
the sample by using the SExtractor morphological classifier.
We defined as stars those sources brighter than I$_{814}$=22 and having a 
value of the ``stellarity'' index larger than 0.9.
This choice tends to underestimate stars both at faint magnitudes where
no classification is considered, and  at bright magnitudes where some fuzzy
stars could be misclassified as galaxy.
On the other hand this will ensure that our galaxy sample is not biased 
against compact galaxies.
The star ``cleaning'' procedure has classified and removed 14 stars 
at I$_{814}<22$  in  agreement with the number of stars found in the HDF-N
by Mendez et al. (1998) to this depth  and in excess by a factor 
of two with respect to the prediction of the galaxy model of 
Bahcall \& Soneira (1981).

\section{Completeness Correction}
Completeness~~ correction~~  for~ faint~ undetected~ sources strongly depends on 
the source apparent spatial structure besides to their magnitude. To estimate completeness via simulations we must account for  sub-galactic structures. These could be quite common at high redshift and detectable by HST, and for the morphology which is that of the co-moving UV and B pass-bands and hence is strongly affected  by star formation episodes.
These features imply that  ``typical'' profiles of galaxies are not able to
well describe the shapes of a lot of galaxies in the HDF-S.  
Thus, in order  to reproduce the manifold of shapes which characterizes
sources in the HDF-S  we generated a set of simulated frames by directly dimming the original frames themselves by various factors while keeping constant the RMS (Sa\-rac\-co et al., 2000).
This procedure has allowed us to avoid any assumption on the source profile
while providing an artificial fair dimmed sample in a real background noise.
We thus define the correction factor $\bar c$ as the mean number of dimmed galaxies which should enter the fainter magnitude bin over the mean number of detected ones. It represents the inverse of the fraction of galaxies undetected in each bin. If  $n_i$ is the number of galaxies detected in the i-th bin of the original catalogue and  $m_{i+1}$ is the number of sources in the  i+1-th bin of the simulated catalogue, the correction factor  $c_{i+1}$ corresponding to the ratio between the expected number of galaxies ($n_i$) and the number of galaxies recovered ( $m_{i+1}$). The ``true'' number of galaxies in the  i+1-th bin of the original catalogue is then  $N_{i+1}$= $n_{i+1}\cdot \bar c_{i+1}$, where $\bar c$ is the mean over different simulations for the same frame.

When dimming fluxes by a factor F=$10^{0.4y}$, magnitudes are $y$ mag fainter, but also noise is lowered: if $\sigma$ is the sky RMS on original images, the dimmed frames have a RMS of  $\sigma /F$. We then added a frame of pure Poissonian noise, with a sky RMS  \[\sigma_{noise}^2= \sigma ^2 -(\sigma /F)^2= \sigma ^2(1- \frac{1}{F^2}). \] The final images have, by construction, the correct RMS. By choosing  $y$=0.5 mag, the artificial noise added is $\sigma_{noise}=0.6\sigma$, i.e. about one third of the final image is due to pure Poissonian noise. As showed in Figures 5-6 and in Table 4, this choice allowed us to correct up to V$_{606}$=29, with brighter limits in the other bands.  

We tried also $y$=1 mag, but the estimated incompleteness was catastrophic, corresponding to 96$\%$. This result seemed caused by the high simulated noise ( $\sigma_{noise}$=0.84 $\sigma$, i.e. almost one half of noise is artificial), which caused also brighter bins to be incomplete.

The completeness correction allowed the computation of differential number counts up to fainter magnitudes: we thus determined the slopes of the number counts relation. Our best fit gives $\gamma_U=0.47\pm0.05$, $\gamma_B\sim0.35\pm0.02$, 
$\gamma_V\sim0.28\pm0.01$ and $\gamma_I\sim0.28\pm0.01$ (see Volonteri et al. 2000, for a detailed discussion).

\begin{table}
\caption{Correction factor  $\bar c$ and error. the factor $\bar c$ accounts for the galaxies missed in the detection due to the influence of noise. The ``true'' (corrected) number of galaxies in each bin is $N=\bar cn$, where $n$ is the raw number of detections.}
\begin{tabular}{cccc}\\
 \hline
Filter 	    & mag &  $\bar c$  & $\sigma_c$   \\  
\hline
F300W       & 26.75 & 2.12  & 0.07 \\
F450W       & 27.25 & 1.11 & 0.05 \\
	    & 27.75 & 1.56 & 0.04\\
            & 28.25& 3.25 & 0.11\\
F606W       & 28.25 & 1.34  & 0.02 \\
	    & 28.75 & 2.67 & 0.1 \\
F814W       & 27.25 & 1.09  & 0.04  \\
  	    & 27.75 & 1.81 & 0.11 \\
F110W       & 27.25 & 1.24  & 0.14 \\
  	    & 27.75 & 2.56 & 0.19 \\
F160W       & 27.25 & 1.93  & 0.42 \\
F222M       & 24.25 & 4.09 & 1.23  \\  
\hline
\end{tabular}
\end{table}

\begin{figure}
\centerline{\psfig{figure=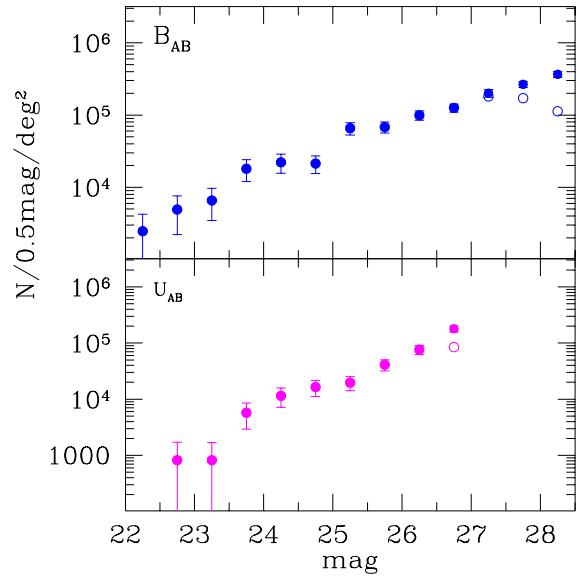,height=80mm}}
\caption{V$_{606}$ and I$_{814}$ number counts: raw counts are shown with empty symbols, corrected counts with filled symbols.}
\end{figure}
\begin{figure}
\centerline{\psfig{figure=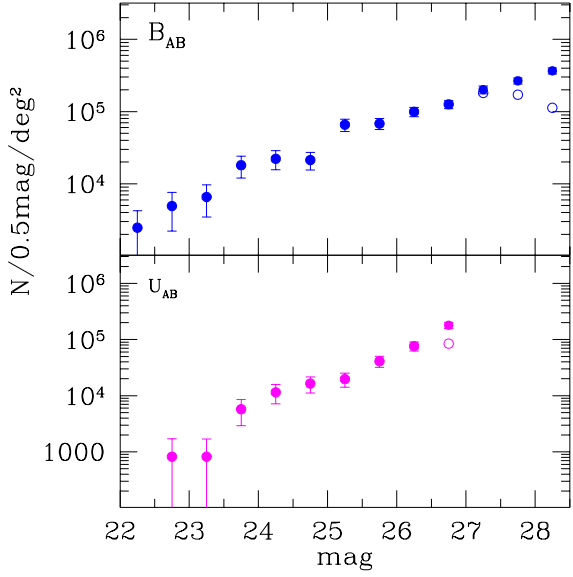,height=80mm}}
\caption{B$_{450}$ and U$_{300}$ number counts: raw counts are shown with empty symbols, corrected counts with filled symbols.}
\end{figure}

\section{Galaxies or HII Regions?}
Colley \etal (1996) suggested that galaxies in the HDF-N  may suffer from a wrong selection. High redshift galaxies on optical images have a clumpy appearance: first the redshift moves the ultraviolet rest-frame light into the optical, so galaxies are observed in UV rest-frame where star-forming regions are more prominent, second the fraction of irregular galaxies is higher than locally (van den Bergh \etal 1996, Abraham \etal 1996) and a large number of galaxies display  asymmetry and multiple structure.

Colley \etal (1996) stressed that compact high-redshift objects may appear more prominently than diffuse objects if their angular size is smaller than the point-spread function (PSF).  The cosmological dimming in surface brightness for these sources is less significant, leading to an enhancement of compact sources over diffuse, resolved objects.  Since HST images have an excellent seeing, these clumps are not~smoothed, so they may confuse detection algorithms  (SExtractor, DAOFIND, FOCAS, etc), and be counted as several distinct faint sources. 

As suggested by Colley \etal (1996) in this case strong correlations between sources on scales $<10$ kpc should be found. A good test is therefore the two-point angular correlation function. 

In order to check our sample against this effect, we computed this function on small angular scales: HII regions physical sizes are  about 0.5 kpc, so a wrong selection of the catalogue should bring a positive peak in the two-point angular correlation function at $0.25-1$ arcsec. This scale corresponds to sizes less than 10 kpc for a wide range in redshift ($0.8 < z < 3.5$).   

We computed the correlation function by comparing the number of data pairs at given angular separation to the number of data-random simulated pairs at the same separation (Davis \& Peebles, 1983). At fixed  $\theta$ the sum of data pairs is \[DD(\theta)=\sum_i\sum_j \delta_i\delta_j\] where $\delta_i$ e $\delta_j$ are delta-functions on i-th and j-th galaxies positions. The sum over $i$ is over all sources in the sample, and the sum over $j$ includes only objects within a distance $\theta$ from particle $i$.
 
We then created a random sample and computed the cross count sum \[DR(\theta)=\sum_i\sum_j \delta_i\delta^R_j\] where $\delta_i$ is as before, and  $\delta^R_j$ is the delta-function for positions of objects in the random sample within a distance $\theta$ from particle $i$.

The resulting  two-point angular correlation function is given by:
\[w(\theta)=\frac{n_R}{n_D}\frac{DD(\theta)}{DR(\theta)}-1\]
where $\frac{n_R}{n_D}$ is the ratio of the mean density of random and data samples respectively. 

We analyzed our sample in I$_{814}$ and B$_{450}$ bands. The I$_{814}$-band catalogue should suffer less from the effects described above, being selected in the reddest filter, the vice versa is true for the B$_{450}$-band catalogue. 

The  two-point angular correlation function is compatible with zero in each band if computed on the whole catalogue. Following Colley \etal (1996) we selected ``small \gxs'' as defined by $\mathcal{D}=\sqrt{ab}<0.2$ arcsec, where $a$, $b$ are the intensity-weighted second moments. At $\theta=0.8$ arcsec it is evident a peak in the function,  $w(\theta)=0.67 \pm 0.59$ (Figure 7). This feature may not be significant due to large errors.  

However about  20-30$\%$ of sources in the B$_{450}$-band catalogue have separation$<1$ arcsec. We therefore analyzed these sources, by cross-correlating  I$_{814}$ and B$_{450}$ catalogues. In the B$_{450}$-band catalogue we selected pairs with a separation $<1$ arcsec which were not included in the I$_{814}$-band catalogue. These objects were single sources splitted in the B$_{450}$-band (with B$_{450} \approx $ 27-29), corresponding to a single detection in the I$_{814}$-band. We then used SExtractor on the B$_{450}$ frame, after choosing a higher  \emph{deblend-mincont}=0.1.  87 sources, with $21<B_{450}<26$, corresponding to about 7$\%$ of the whole sample, were then  considered as single galaxies.

\begin{figure}
\centerline{\psfig{figure=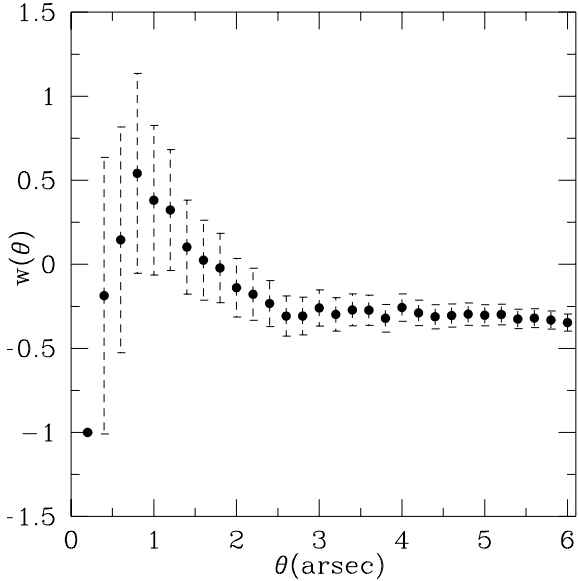,height=80mm}}
\caption{The two-point angular correlation function $w(\theta)$ estimated in the B$_{450}$ band on small angular scales, for ``small'' sources, as defined by $\mathcal{D}=\sqrt{ab}<0.2$ arcsec, where $a$, $b$ are the intensity-weighted second moments. The function is compatible with zero, but, within the error bars, it is possible to notice a peak at $\theta=0.8$ arcsec (corresponding to a physical distance of 4-6 kpc for 0.3$<z<$5). }
\end{figure}

This simple analysis shows that a remarkable fraction of sources in the HDF-S has a  neighbour at small angular distance. It should be pointed out that the HDF-S is a pencil-beam survey, and projection effects may be sizeable.

\section{Optical Colours}
The estimate of galaxy colours could be biased towards the selection band and affected by an aperture effect. In order to test for this, we first determined colours simply by using SExtractor in the so-called \emph{double image mode} on the I$_{814}$-selected catalogue, we measured B$_{450}$ magnitudes within the 1$\sigma$ isophotal area determined in the I$_{814}$ band (B$_{col}$). The (B-I)$_I$ colour is then the difference between B$_{col}$ and the 1$\sigma$ isophotal magnitude in the I$_{814}$ band. We then measured the (B-I)$_B$ colour for the B-selected catalogue using the same method, but with the B$_{450}$ band  1$\sigma$ isophotal area as a reference for both the I$_{814}$  (I$_{col}$) and B$_{450}$ magnitudes. 

For every source the B-I colours measured on the basis of the B$_{450}$ or I$_{814}$ area are different. In Figure 8 we show the residuals  (B-I)$_I$-(B-I)$_B$ vs I$_{814}$. (B-I)$_B$ is slightly redder than (B-I)$_I$),~ probably~  because ~the ~B$_{450}$-band ~ isophote~underestimates the I$_{814}$ band flux (broadly speaking I$_{814}$ isophotal areas are larger than B$_{450}$ isophotal areas).

In order to measure colours unbiased by the selection band we chose a different approach, that is we  created a \emph{meta-image} UBVI by summing all four frames, after normalizing each to have the same rms sky noise. We then run SExtractor in the \emph{double image mode}: detection and isophote boundaries were measured on the combined image, while isophotal magnitudes were measured on U$_{300}$, B$_{450}$, V$_{606}$, I$_{814}$ images individually. Using this procedure (Moustakas et al. 1997) both object detection and isophote determination are based on the summed image, and isophotes are not biased towards any of the bands.

We finally  cross-correlated the catalogue obtained from the combined image with the I$_{814}$ sample selected according to our criteria (see \emph{Image Analysis}). We  assigned a lower limit in magnitude to sources undetected in any of other bands (that is missing in our final  U$_{300}$, B$_{450}$, V$_{606}$ catalogues). The limiting magnitude is the 5$\sigma$ isophotal magnitude within the isophote measured in the combined image.

\begin{figure}
\centerline{\psfig{figure=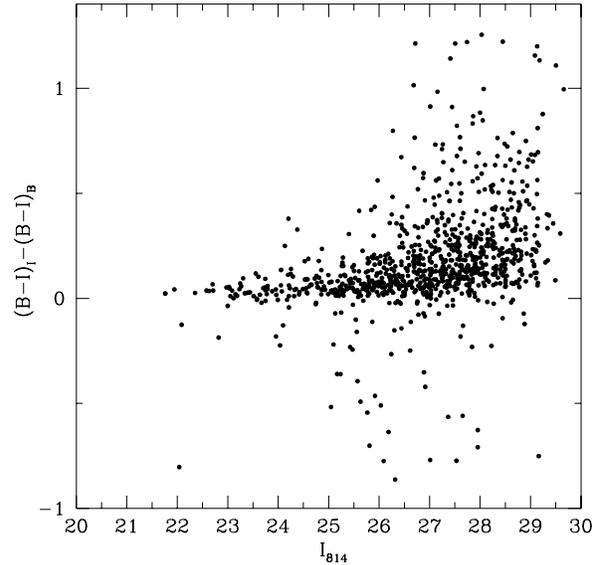,height=80mm}}
\caption{Difference between (B-I) computed in I$_{814}$ or B$_{450}$ bands vs I$_{814}$. Colours are measured having as a reference the  1$\sigma$ I$_{814}$ and B$_{450}$  isophotal areas respectively. (B-I)$_I$-(B-I)$_B$ correlates with the I$_{814}$ magnitude: (B-I)$_B$ is slightly redder than (B-I)$_I$ at faint magnitudes, probably because the B$_{450}$-band isophote underestimates the I$_{814}$ band flux.}
\end{figure}

\section{Structural Parameters}
For a subsample (I$_{814}$$<26$) we computed size, surface brightness, light concentration index, asymmetry index.


The measure of  the size of distant galaxies needs the use of a metric radius, that is a size independent on the redshift and light profile. We estimated  a metric radius based on the Petrosian function (Petrosian, 1976), defined (Kron, 1995) as  \[\eta(\theta)=\frac{1}{2}\frac{d\ln l(\theta)}{d \ln(\theta)}\] where $l(\theta)$ is the light growth curve. This function has the property: \[\eta(\theta)=\frac{I(\theta)}{<I>_{\theta}}\] where $I(\theta)$ is the surface brightness at the radius $\theta$ and $<I>_{\theta}$ is the mean surface brightness within $\theta$. We defined for each galaxy the angular size as the value of $\theta_{0.5}$ such that $\eta(\theta)=$0.5 (Bershady et al., 1998, Saracco et al, 1999). In order to determine the function $\eta(\theta)$ we obtained the intensity profile, through multi-aperture photometry at equispaced (0.04 pixel) diameters, and subsequently interpolated it by a spline fit. There are sources whose $\eta(\theta)$ is always larger than 0.5 (considering as a border the aperture $i-th$ such that mag$_{i+1}>$  mag$_{i}$); in these cases we defined  $\theta_{0.5}=-1$, and also all quantities linked to  $\theta_{0.5}$ were arbitrarily set $-1$. We also measured the effective radius (half-light radius, $r_{eff}$); for I$_{814}$$<26$ the relation $\theta$ vs $r_{eff}$ is well fitted by $\theta_{0.5}=1.2r_{eff}$. After determining $\theta_{0.5}$, we computed for each galaxy the mean surface brightness within $\theta_{0.5}$.

Abraham~ et al. (1994) ~and ~Abraham et al. (1996) showed that two indexes, namely an asymmetry index ($A$) and a central concentration index ($C_{abr}$), are very useful in order to estimate a quantitative galaxy morphology. The former is determined by rotating the galaxy by 180$^\circ$ and subtracting the resulting image from the original one. The asymmetry index is given by the sum of absolute values of the pixels in the residual image, normalized by the sum of the absolute value of the pixels in the original image and  corrected for the intrinsic asymmetry of the background. The concentration index is given by the ratio of fluxes in two isophotes, based on the analysis of light profiles. The measure of these indexes is independent of colour, though they correlate well with optical colours. 

We therefore computed both the asymmetry and the concentration indexes following Abraham et al. (1994) and Abraham et al. (1996), but also computed a different light concentration parameters.

That is we  computed \[C_\eta=\frac{F(<\theta_{0.5})}{F(<1.5\theta_{0.5})}\] (Saracco et al., 1999), i.e. the ratio between the flux within radius $\theta_{0.5}$ and the flux within $1.5\theta_{0.5}$.

As discussed by Saracco et al. (1999), $C_\eta$ is independent from the redshift of the source, since it is related to a metric size and is independent of the asymptotic profile. Brinchmann et al. (1998) on the contrary pointed out that the central concentration $C_{abr}$ defined by Abraham et al. (1994, 1996) is redshift-dependent.  $C_{abr}$ has been computed in order to classify our galaxies and compare our results with HDF-North. 
\begin{figure}
\centerline{\psfig{figure=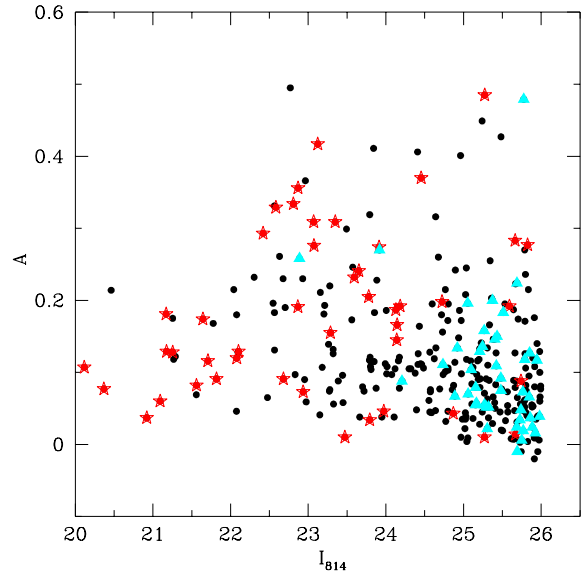,height=80mm}}
\caption{Asymmetry index vs I$_{814}$ magnitude: sources with both B$_{450}$-V$_{606}$ and V$_{606}$-I$_{814}$ redder than local ellipticals are shown as stars, while sources bluer than local irregulars are shown as triangles.}
\end{figure}
We tested the various parameters against apparent magnitude and checked any correlation with colours (Conselice et al. 2000). We selected two subsamples, composed respectively by galaxies with both B$_{450}$-V$_{606}$ and V$_{606}$-I$_{814}$ redder or bluer than a local elliptical or a local irregular galaxy. We used these subsamples as tracers of colours.   

The asymmetry index (Figure 9) seems not to be biased: the faintest sources are on the whole more symmetric than brighter ones, but the presence of asymmetric objects also in the last bins suggest this feature to be linked to the nature of these galaxies. The trend towards high symmetry may be due to the influence of noise, which makes the profile smoother.

The central concentration index $C_{abr}$ defined by Abraham et al. (1994, 1996) seems to be biased against compact sources at faint magnitudes (I$_{814}>24.5$), while  $C_\eta$ does not correlate with apparent magnitude (Figures 10-11)
\begin{figure}
\centerline{\psfig{figure=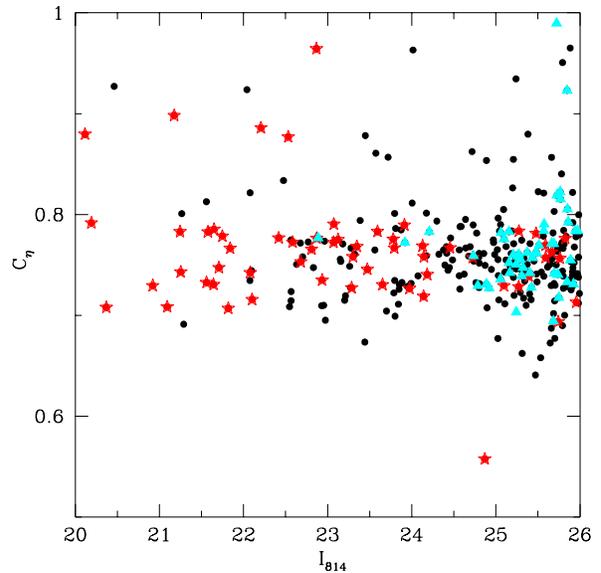,height=80mm}}
\caption{Central concentration index $C_\eta$  vs I$_{814}$ magnitude: sources with both B$_{450}$-V$_{606}$ and V$_{606}$-I$_{814}$ redder than local ellipticals are shown as stars, while sources bluer than local irregulars are shown as triangles.}
\end{figure}
\begin{figure}
\centerline{\psfig{figure=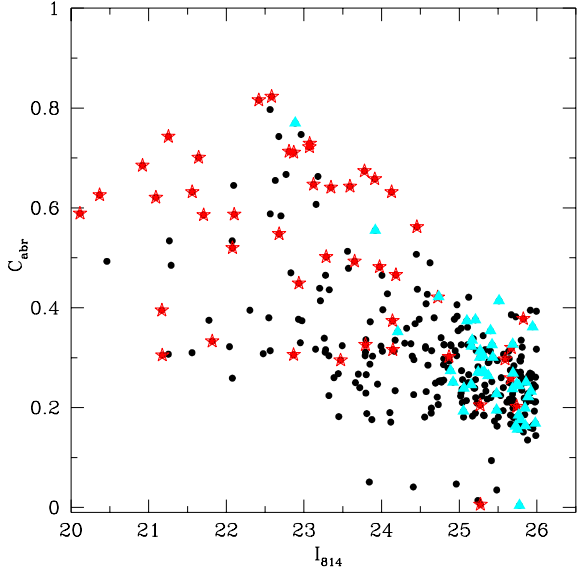,height=80mm}}
\caption{Central concentration index $C_{abr}$  vs I$_{814}$ magnitude: sources with both B$_{450}$-V$_{606}$ and V$_{606}$-I$_{814}$ redder than local ellipticals are shown as stars, while sources bluer than local irregulars are shown as triangles.}
\end{figure}

\section{Description of the Catalogue}
We presented a catalogue of galaxies in the HDF-S, created using the public version 1 images of the WFPC2 data. We created a catalogue for each pass-band (I$_{814}$, V$_{606}$, B$_{450}$$_{450}$, U$_{300}$), which is available upon request. In  V$_{606}$, B$_{450}$, U$_{300}$ for each galaxy the catalogue reads:
\begin{itemize}
\item isophotal flux and magnitude, with their errors (SExtractor)
\item  isophotal corrected flux and magnitude, with their errors (SExtractor)
\item  fixed circular aperture (0.6 arcsec radius) flux and magnitude, with their errors (SExtractor) 
\item  Kron's technique flux and magnitude, with their errors (SExtractor)
\item SExtractor best choice for photometry (Kron's or isophotal corrected)
\item  isophotal area (pix$^2$)(SExtractor)
\item coordinates ~~in ~pixels ~of ~the barycentre,~related to version1 frames (SExtractor)
\item  right ascension and declination (SExtractor)   
\item second order moments of the light distribution (SExtractor) 
\item semi-major and semi-minor axis lenghts in pixels (SExtractor)
\item flags produced by detection and measurement processes (SExtractor)
\item stellar classification (SExtractor) 
\item signal-to-noise ratio (Section 4)
\item pseudo-total magnitude and error (Section 3)
\end{itemize}

For the I$_{814}$-selected catalogue besides the above entries, we estimated also the colours (Section 7) and for I$_{814}<26$, the petrosian radius (arcsec), the mean surface brightness within the petrosian radius (mag arcsec$^{-2}$), light concentration indexes, that is $C_\eta$ and $C_{abr}$, and the  asymmetry index as computed by Abraham software (Section 8).
In Table 5 a small part of the catalogue is shown. 

\appendix
\section{The Catalogue}
Example of a part of the catalogue: we show only some photometric entries for sake of concision. The real catalogue reads also the  isophotal corrected flux and magnitude, with their errors, a fixed circular aperture (0.6 arcsec radius) flux and magnitude, with their errors, Kron's technique flux and magnitude, with their errors, and SExtractor best choice for photometry (Kron's or isophotal corrected). The full catalogue is available at http://cdsweb.u-strsbg.fr/Abstract.html and via \\
\noindent
http://www.merate.mi.astro.it/~saracco/science.html.
\begin{table*}[!h]
\begin{tabular}{ccccccc}
\\ 
\hline
FLUX-ISO 
& FLUXERR-ISO
&MAG-ISO
&MAGERR-ISO 
&MTOT 
&MTOT-ERR
&ISOAREA-IMAGE

\\
\hline		
    0.354 &   0.0012  &    23.21 &  0.0037  &    23.17 &    0.1 &    1027 \\
    0.037 &  7.9E-4  &    25.64  &  0.022 &    25.61 &    0.1   &    338 \\
   0.024 &  4.3E-4  &    26.11  &  0.019 &    25.81  &   0.1   &    139  \\
    0.111 &  7.8E-4  &    24.47  & 0.007  &    24.32 &    0.1  &     470 \\
    0.035 &  6.2E-4  &    25.70 &   0.019  &    25.47 &    0.1   &    279 \\
   0.027 &  5.7E-4  &     25.99 &   0.023  &    25.78 &    0.1   &    236\\
    0.125 &  8.0E-4  &    24.34  & 0.007  &    24.20 &    0.1  &    507 \\
 \hline
\end{tabular}

\begin{tabular}{ccccccc}
\\ 
\hline
X-IMAGE
&Y-IMAGE
&ALPHA-J2000 
&DELTA-J2000
&X2-IMAGE
&Y2-IMAGE
&ERRX2-IMAGE 
\\ 
\hline
   270.59  &    1900.766  &    338.272981  &   -60.555908 &     47.9 &   25.41 &  0.0016 \\ 
     290.17   &   1569.351  &    338.272480   &  -60.559576 &     44.6 &    29.47 &   0.0252\\
   304.80   &   1706.001  &    338.272175   &  -60.558062  &    8.2  &   11.51  &0.0044  \\
      321.70   &   1995.368  &    338.271847   &  -60.554856 &     26.4 &    20.98 &  0.0023\\ 
      372.10   &   1877.157   &   338.270679  &   -60.556161  &    27.9 &  15.58   & 0.0189 \\
    375.62 &    717.7169  &    338.270398   &  -60.568997  &    31.6 &   20.41  & 0.0026\\
 \hline
\end{tabular}

\begin{tabular}{ccccccc}
\\ 
\hline
ERRY2-IMAGE
&A-IMAGE
&B-IMAGE 
&ERRA-IMAGE
&ERRB-IMAGE
&FLAGS
&CLASS-STAR\\ 
\hline
 7.9E-4  &    7.08 &  4.81 &   0.041  &  0.02672239 &  0 &    0.91  \\
    0.018  &    6.69 &  5.40 &    0.15  &   0.1358412 &  2 &  2.2E-4  \\
 0.006   &      3.44  &  2.81 &  0.079   & 0.06579173  & 0  &  0.02    \\
  0.001 &    5.33  &   4.35 & 0.05 &    0.04017934   &0   & 0.02    \\
 0.010  &    5.34  &   3.87 &   0.14  & 0.09491127   &0   &3.0E-4  \\
 0.0015  &     5.63  &     4.50&   0.05  & 0.03895507   &0   & 0.02  \\
 \hline
\end{tabular}

\begin{tabular}{ccccccc}
\\ 
\hline
SN
&B-V
&THETA
&MUTHETA
&CM 
&C-ABR
\\ 
\hline
   133.95 &   0.53  &            0.1489   &      21.11 &       0.7708    &     0.663\\
    25.60 &    0.43 &               0.4587  &        25.5&        0.7479   &      0.211\\
  25.55   &-0.07    &         0.1899     &    24.12   &     0.7705      &   0.321\\
 62.89    &0.52    &         0.2793       &  23.38     &   0.7606        & 0.359\\
   22.07   &   0.67    &           0.3213   &      25.08 &       0.7711     &    0.157\\
   68.09    & 0.10    &         0.3192     &    23.46   &     0.7829       &  0.352\\

\hline
\end{tabular}
\end{table*}

\begin{acknowledgements}
Thanks to M. Bolzonella for useful discussions and support and R. Abraham for making available the software for computing asymmetry and concentration index. We would like to thank the referee, G. Paturel, for his helpful comments and suggestion. MV acknowledges financial support by Fondazione Cariplo. 

\end{acknowledgements}

\end{document}